\documentclass[conference, a4paper]{IEEEtran}
\IEEEoverridecommandlockouts

\usepackage{graphicx}
\usepackage{url}
\usepackage{amsmath}
\usepackage{multirow}
\usepackage{bm}
\usepackage{multicol}
\usepackage[hidelinks]{hyperref}
\usepackage{cleveref}
\usepackage{amssymb}

\def\BibTeX{{\rm B\kern-.05em{\sc i\kern-.025em b}\kern-.08em
    T\kern-.1667em\lower.7ex\hbox{E}\kern-.125emX}}

\DeclareRobustCommand*{\IEEEauthorrefmark}[1]{%
    \raisebox{0pt}[0pt][0pt]{\textsuperscript{\footnotesize\ensuremath{#1}}}}

\begin{document}

\title{Unsupervised Tumor-Aware Distillation for Multi-Modal Brain Image Translation}

\author{
\IEEEauthorblockN{
Chuan Huang\IEEEauthorrefmark{1},
Jia Wei\IEEEauthorrefmark{1}*\thanks{* Jia Wei is the corresponding author.},
Rui Li\IEEEauthorrefmark{2}}

\IEEEauthorblockA{\IEEEauthorrefmark{1}School of Computer Science and Engineering, South China University of Technology, Guangzhou, China\\ 
\href{mailto:cschuan@mail.scut.edu.cn}{cschuan@mail.scut.edu.cn} \qquad
\href{mailto:csjwei@scut.edu.cn}{csjwei@scut.edu.cn}
\IEEEauthorblockA{\IEEEauthorrefmark{2}Golisano College of Computing and Information Sciences, Rochester Institute of Technology, Rochester, NY, USA \\ \href{mailto:rxlics@rit.edu}{rxlics@rit.edu}}
}
}

\maketitle

\begin{abstract}
Multi-modal brain images from MRI scans are widely used in clinical diagnosis to provide complementary information from different modalities. However, obtaining fully paired multi-modal images in practice is challenging due to various factors, such as time, cost, and artifacts, resulting in modality-missing brain images. To address this problem, unsupervised multi-modal brain image translation has been extensively studied. Existing methods suffer from the problem of brain tumor deformation during translation, as they fail to focus on the tumor areas when translating the whole images. In this paper, we propose an unsupervised tumor-aware distillation teacher-student network called UTAD-Net, which is capable of perceiving and translating tumor areas precisely. Specifically, our model consists of two parts: a teacher network and a student network. The teacher network learns an end-to-end mapping from source to target modality using unpaired images and corresponding tumor masks first. Then, the translation knowledge is distilled into the student network, enabling it to generate more realistic tumor areas and whole images without masks. Experiments show that our model achieves competitive performance on both quantitative and qualitative evaluations of image quality compared with state-of-the-art methods. Furthermore, we demonstrate the effectiveness of the generated images on downstream segmentation tasks. Our code is available at \url{https://github.com/scut-HC/UTAD-Net}.
\end{abstract}


\section{Introduction}
\begin{figure}
    \centering
    \includegraphics[width=1.0\linewidth]{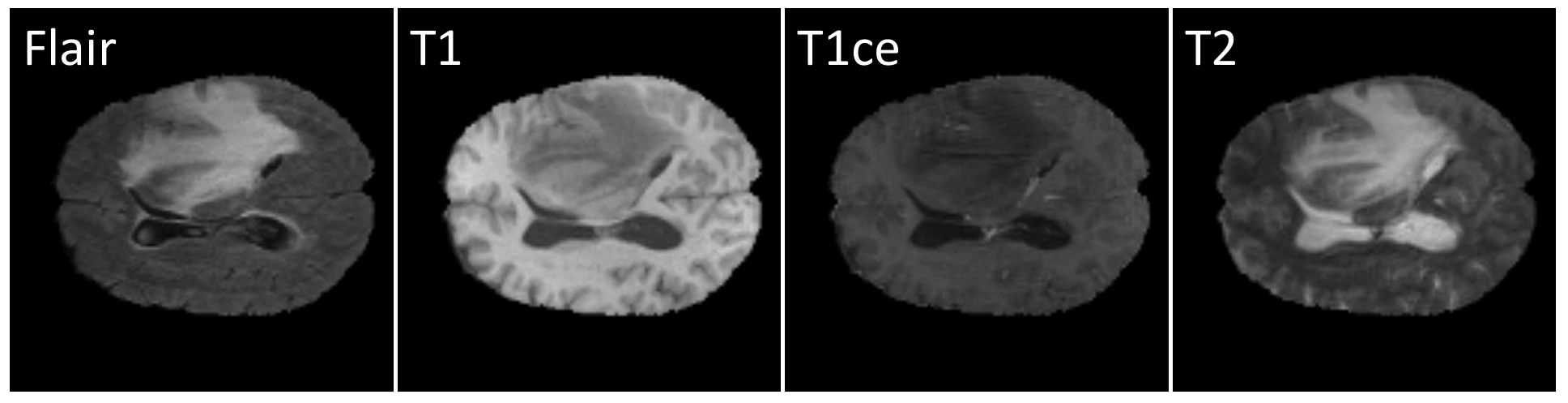}
    \caption{Examples of brain tumor MRI images from different modalities.}
    \label{fig1}
\end{figure}

Multi-modal brain images from MRI (Magnetic Resonance Imaging) scans are widely used in various clinical scenarios\cite{icsin2016review,hou2021brain}. These images are further divided into several modalities(sequences), such as T1-weighted (T1), T1-with-contrast-enhanced (T1ce), T2-weighted (T2), T2-fluid-attenuated inversion recovery (Flair), etc. Each modality exhibits distinct contrasts, providing complementary lesion information from different perspectives. As shown in \Cref{fig1}, Flair and T2 images depict the peritumoral edema areas clearly, while T1 images highlight the white and gray matter tissues\cite{liu2020multimodal}, making them suitable for presenting anatomical structures. T1ce images delineate the structures and the edges of the tumors\cite{liu2023learning}, which is convenient to observe the morphology of different types of tumors. Fully paired multi-modal images assist doctors in achieving more precise diagnoses\cite{xin2020multi}.

The benefits of using multi-modal images to assist medical analysis have been widely recognized\cite{yuan2020unified}. However, physicians usually obtain some of the modalities in practice due to practical considerations such as time, cost, and artifacts. As a result, many images are modality-missing, which can negatively impact the accuracy of diagnoses. Consequently, there has been significant research attention on generating images of missing modalities\cite{zhou2020brain,zhou2021latent}.

Existing methods for multi-modal image translation have shown promising results in natural images. However, when applied to medical images, particularly brain tumor images, the results are often unsatisfactory\cite{liu2020multimodal}. Compared with two-dimensional natural images, three-dimensional medical images have more structural information\cite{zhou2021models}, leading to blur or deformation in image translation. Moreover, due to the privacy of patients, a large number of medical images collected by different institutions are private, which increases the difficulty of model training. Therefore, translating brain tumor images remains a challenging task. Some methods propose using segmentation or object detection networks to assist translation\cite{shen2019towards,chen2021targan,bhattacharjee2020dunit}, while others propose loss functions to preserve tumor information\cite{zhang2018translating,xin2020multi, huang2022ds}. However, they either require paired images for training or still rely on additional networks or labels during inference.

To solve the problem of brain tumor deformation during translation, we propose an Unsupervised Tumor-Aware Distillation Teacher-student Network called UTAD-Net. Our model comprises a teacher network and a student network. The teacher network consists of two branches: the global branch translates the input whole images into the target modality, while the local branch translates the tumor images obtained through pixel-level multiplication of the whole images and their corresponding tumor masks. Each branch is composed of a pair of encoders and decoders. Unlike other distillation models, the student network adopts an identical network structure as the teacher network but reduces the input information. What is inputted into the local branch is not the tumor images, but the whole images. The encoder of the local branch in the student network is trained to perceive the tumor areas and learn features similar to those encoded by the encoder of the teacher network for the tumor images. We also propose a local consistency loss to preserve the anatomical structure of the tumors. All training samples consist of unpaired single-modality images. In this way, the student network can focus on the tumor areas and generate more realistic images without the tumor masks. During the inference phase, we directly employ the student network to translate the input source modality images into the final target modality images.

In summary, the main contributions of this paper are as follows:
\begin{itemize}
    \item we propose UTAD-Net to accurately perceive the tumor areas and generate more realistic images with clearer textures and richer structural details using unpaired brain tumor images, assisted by additional tumor masks and the local consistency loss. 
    \item To the best of our knowledge, we are the first to propose reducing the input information of the student network in the distillation model. This approach allows the teacher network to capture the tumor areas with the guidance of the tumor masks, while the student network acquires knowledge from the teacher network through distillation and gains the ability to generate images without masks.
    \item We present qualitative and quantitative evaluations of image quality for multi-modal brain images translation on the BRATS2020 dataset\cite{bakas2018identifying}. Our model outperforms state-of-the-art methods. We also demonstrate that the images generated by our model are helpful for improving the segmentation performance through downstream tasks.
 
\end{itemize}

\section{Related works}
Cross-modality image translation has been intensively studied in recent years. For instance, Pix2pix\cite{isola2017image} presents a cGAN-based\cite{mirza2014conditional} method to convert images between source and target modalities. However, it requires paired data for training, which is hard to realize. Therefore,  unsupervised image translation using unpaired data has piqued researcher interest. CycleGAN\cite{zhu2017unpaired} and DiscoGAN\cite{kim2017learning} propose a cycle consistency loss, which attempts to preserve the crucial information of the images. By constraining the reconstructed image and the source image, the model is available to translate images between the two given modalities with unpaired data. UNIT\cite{liu2017unsupervised} assumes that the two modalities share the same latent space, and proposes to combine VAE and GAN to form a more robust generative model. The encoder maps the images of different domains to the same distribution to obtain the latent code, and then the decoder maps the latent code back to the image domain. However, these models fail to generate diverse styles in the images. The resulting target modal image for a given input image is singular. In order to solve this problem, MUNIT\cite{huang2018multimodal} and DRIT\cite{lee2018diverse} disentangle the latent code into the content code which is shared by different modalities and the style code which is unique for different modalities and restricted to normal distribution. During inference, the style code can be obtained through sampling, allowing the model to generate images with diverse styles.

However, the above models can only translate images between two modalities. If we want to translate images between $n$ modalities, we need to train the model for $n(n-1)/2$ times. In order to perform in a unified model to translate multi-modality images, StarGAN\cite{choi2018stargan} proposed a single generator to learn the mapping between any two given modalities. The source images and mask vectors are inputted to the generator which then outputs the generated target images. The discriminator needs to not only distinguish whether the images are real or false, but also classified the domain they belong to. Since every mask vector is corresponding to a given condition, the generated images are simplex without style diversity. DRTI++\cite{lee2020drit++} adds domain codes for translation so that any target modality images can be generated by a unified generator. ResViT\cite{dalmaz2022resvit} introduces a transformer-based model that exhibits broad adaptability across diverse configurations of source-target modalities.

Although the above model can achieve multi-modality translation, it can not focus on local targets but only on the whole image. CSCG\cite{zhang2018translating} propose that for unsupervised learning, cycle consistency loss will easily lead to local deformation of the image if there are no other constraints. InstaGAN\cite{mo2018instagan} proposed to add segmentation labels of local instances as additional input information so that the network will pay more attention to the shape of the local instances in the training process and reduces the deformation. DUNIT\cite{bhattacharjee2020dunit} and INIT\cite{shen2019towards} respectively propose to use object detection and segmentation to assist translation. Ea-GANs\cite{yu2019ea} proposes to integrate edge maps that contain critical textural information to boost synthesis quality. TC-MGAN\cite{xin2020multi} introduces a multi-modality tumor consistency loss to preserve the critical tumor information in the target-generated images but it can only translate the images from the T2 modality to other MR modalities. TarGAN\cite{chen2021targan} focuses on the target area by using a shape controller. While these models can translate images more effectively, they also require more supervised information.

Some of the above methods can only translate images between two given modalities, and some require paired and labeled data for training, which is not completely consistent with the practical application scenarios that most data are unpaired. We propose UTAD-Net to learn an end-to-end mapping from an arbitrary source modality to the given target modality, which can focus on the local tumor areas and translate better by using unpaired images. 

\begin{figure*}[t]
\centering
\includegraphics[width=1.0\textwidth]{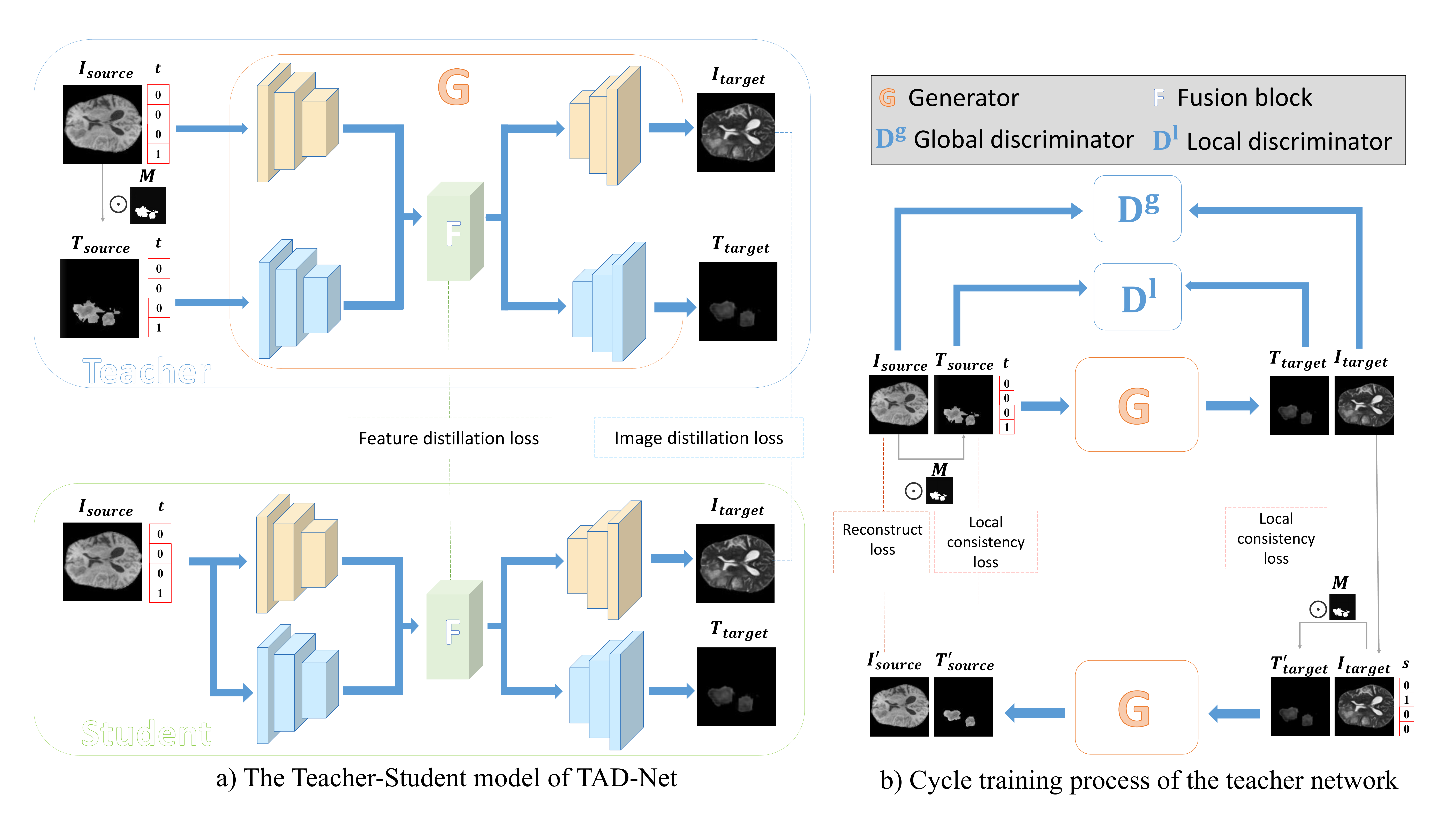} 
\caption{Overview of the proposed UTAD-Net. a) The generator $G$ comprises two encoder-decoder pairs and a fusion block $F$ to generate the whole image and tumor image. The student network learns the knowledge of the teacher network through distillation learning at both the feature and image levels. b) The generator $G$ tries to reconstruct the whole image and tumor image. The global discriminator $D^g$ determines whether the whole image is real or fake and classifies its modality, while the local discriminator $D^l$ models the tumor image. }
\label{fig2}
\end{figure*}

\section{Methods}
In this section, we first describe our framework and the pipeline of our approach, then we define the training objective functions.

\subsection{Framework and pipeline}
For the teacher network, given an image $I_{source}$ from the source modality $s$ and corresponding mask $M$, we aim to train a model that can translate both the whole image and the tumor area to the target modality $t$. To achieve this, we first multiply the two on pixel-level to get the source tumor image $T_{source}$. Then the source whole image $I_{source}$ is fed to the global branch to generate the target whole image $I_{target}$, and the source tumor image $T_{source}$ is fed to the local branch to generate the target tumor image $T_{target}$, respectively. The mapping is denoted as: $(I_{target},T_{target})=G(I_{source},T_{source},t)$. 

The student network adopts the identical network structure as the teacher network, but with a key difference: it does not require the tumor image $T_{source}$ as input. Unlike the teacher network, the local branch of the student network takes the whole image as input instead of the tumor image, allowing the model to generate the target image without a tumor mask, which is denoted as: $(I_{target},T_{target})=G(I_{source},t)$. The teacher-student model is shown in \Cref{fig2}.

\textbf{Generator G.} The generator is comprised of two encoder-decoder pairs, one for the global branch and the other for the local branch. We use a fusion block $F$ proposed by Hi-Net\cite{zhou2020hi} to fuse features from the encoders of the two branches. The decoders of the two branches receive the fused feature and respectively generate the target global image $I_{target}$ and the target tumor image $T_{target}$. Then the generator translates the target whole image $I_{target}$ and its corresponding tumor image(only used in the teacher network) $T_{target}'$ to the reconstructed whole image $I_{source}'$ and tumor image $T_{source}'$. In this way, a cycle training process is accomplished. 

\textbf{Discriminator \bm{${\rm D^g}$} and \bm{${\rm D^l}$}.} Similar to StarGAN\cite{choi2018stargan}, the discriminator tries to not only distinguish whether the image is real or fake but also judge the modality to which it belong. Our model utilizes two discriminators: $D^g$ is responsible for the whole image in the global branch and $D^l$ is responsible for the tumor image in the local branch respectively.

\subsection{Training objective functions}
\textbf{Adversarial loss.} Adversarial loss is designed to make the images generated by the generator more realistic to confuse the discriminator. To train the model more stably, we use the adversarial loss WGAN-GP\cite{gulrajani2017improved} as follows:
\begin{align}
\mathcal{L}_{adv}^{g}=&\mathbb{E}_{I_{source}}[D_{src}^g(I_{source})]-\mathbb{E}_{I_{target}}[D_{src}^g(I_{target})] \nonumber\\
&-{\lambda}_{gp}\mathbb{E}_{\hat{I}}[({\Vert}{\nabla} D_{src}^{g}(\hat{I}){\Vert}_2-1)^2], \\
\mathcal{L}_{adv}^{l}=&\mathbb{E}_{T_{source}}[D_{src}^l(T_{source})]-\mathbb{E}_{T_{target}}[D_{src}^l(T_{target})] \nonumber\\
&-{\lambda}_{gp}\mathbb{E}_{\hat{T}}[({\Vert}{\nabla} D_{src}^{l}(\hat{T}){\Vert}_2-1)^2],  
\end{align}
where \^{I} and \^{T} are uniformly sampled along a straight line connecting the corresponding real images and generated images. ${\lambda}_{gp}$ is set as 10.0 in our model. $D_{src}^g$ and $D_{src}^l$ represent the probability distributions of real or fake for the images generated by the global and local branches respectively.

\textbf{Modality classification loss.} Given an image and its target modality vector, we hope the generator can generate images that are as close to the target modality as possible. Similar to StarGAN, the discriminators aim to judge the modality they belong to. For real images, we define the modality classification loss which is used to optimize the discriminators as follows:
\begin{gather}
\mathcal{L}_{r\_cls}^{g}=\mathbb{E}_{I_{source}, s}[-logD_{cls}^g(s|I_{source})], \\
\mathcal{L}_{r\_cls}^{l}=\mathbb{E}_{T_{source}, s}[-logD_{cls}^l(s|T_{source})],
\end{gather}
where the term $D_{cls}^g(s|I_{source})$ and $D_{cls}^l(s|T_{source})$ represent the probability distribution over the modality vector for the whole images and tumor images. Similarly, we define the modality classification loss for fake images which is used to optimize the generator as follows:
\begin{gather}
    \mathcal{L}_{f\_cls}^g=\mathbb{E}_{I_{target},t}[-logD_{cls}^g(t|I_{target})],\\ 
    \mathcal{L}_{f\_cls}^l=\mathbb{E}_{T_{target},t}[-logD_{cls}^l(t|T_{target})], \\ 
    \mathcal{L}_{f\_cls}=\mathcal{L}_{f\_cls}^g+\mathcal{L}_{f\_cls}^l.
\end{gather}

\textbf{Local consistency loss.} We aim to constrain the similarity of the generated tumor images $T_{target}$ and the tumor areas of the generated whole images $T_{target}'$ to alleviate the problem of distortion in brain tumor image translation. The reconstructed tumor images $T_{source}'$ and the source tumor images $T_{source}$ are constrained in the same way. We thus propose a local consistency loss as an extra constraint to improve the translation effect, which is defined as follows:
\begin{align}
    \mathcal{L}_{local}=&\mathbb{E}([{\Vert}T_{target}-T_{target}'{\Vert}_1])\nonumber\\
    &+\mathbb{E}([{\Vert}T_{source}-T_{source}'{\Vert}_1]).
\end{align}

\textbf{Reconstruct loss.} A challenge is to ensure that the generated images $I_{target}$ just simply change the image style information and still contain all the content information of the source images $I_{source}$. To solve this problem, we feed $I_{target}$ into the translation network for cycle translation to obtain the reconstructed images $I_{source}'$. To minimize the difference between  $I_{source}'$ and $I_{source}$, we add a reconstruct loss which is defined as follows:
\begin{equation}
    \mathcal{L}_{rec}=\mathbb{E}[{\Vert}I_{source}-I_{source}'{\Vert}_1].
\end{equation}

\textbf{Distillation loss.} To allow the student network to perceive the tumor areas like the teacher network without masks, we impose constraints on the intermediate layer features and the final generated target images of both networks, so that the student network can better learn the knowledge from the teacher network. The distillation loss is defined as follows:
\begin{align}
    \mathcal{L}_{dis}=&\mathbb{E}[{\Vert}F^{teacher}-F^{student}{\Vert}_1] \nonumber\\
    &+\mathbb{E}[{\Vert}I^{teacher}_{target}-I^{student}_{target}{\Vert}_1],
\end{align}
where $F^{teacher}$ and $F^{student}$ mean the fused feature of the teacher and student networks.

\textbf{Total loss.} Combining all the losses mentioned above, we finally defined the objective function as follows:
\begin{gather}
\mathcal{L}_D^{g/l}=-\mathcal{L}_{adv}^{g/l}+\mathcal{L}_{r\_cls}^{g/l}, \\
\mathcal{L}^{teacher}_G=\mathcal{L}_{adv}^g+\mathcal{L}_{adv}^l+\mathcal{L}_{f\_cls}+{\lambda_1}(\mathcal{L}_{rec}+\mathcal{L}_{local}) \\ 
\mathcal{L}^{student}_G=\mathcal{L}^{teacher}_G+{\lambda_2}\mathcal{L}_{dis},
\end{gather}
where ${\lambda_1}$ and ${\lambda_2}$ are hyper-parameters to balance losses. We set ${\lambda_1}$ and ${\lambda_2}$ to be 10.0 in our experiments.

\begin{figure*}[!t]
\centering
\includegraphics[width=1.0\textwidth]{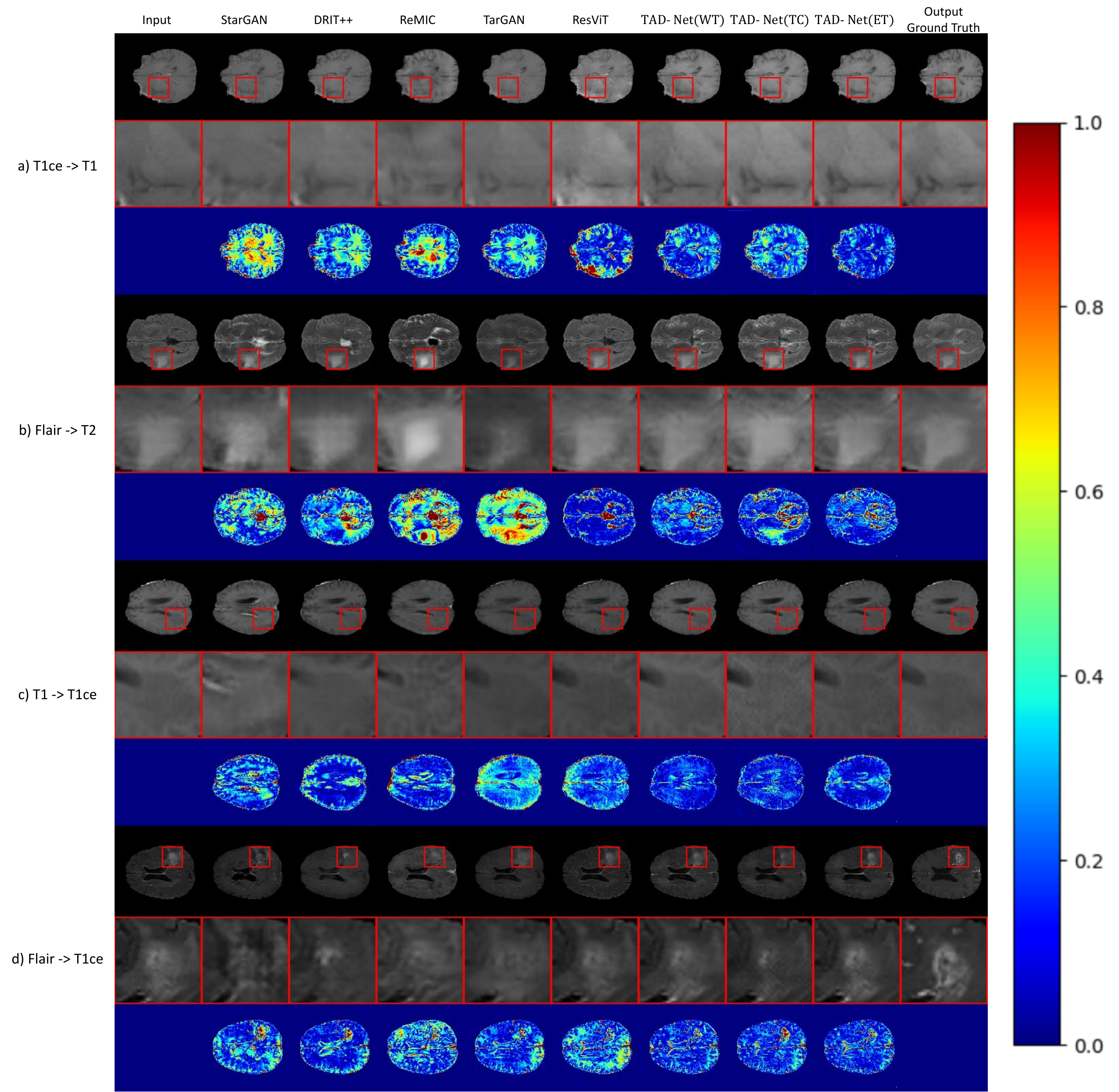} 
\caption{Qualitative evaluations of our model and the other baselines on the BRATS2020 dataset. The four samples are denoted as: a)T1ce$\rightarrow$T1. b)Flair$\rightarrow$T2. c)T1$\rightarrow$T1ce. d)Flair$\rightarrow$T1ce. For each sample, the first row represents the translation of the whole images, the second row depicts the magnified image of the tumor areas, and the third row illustrates the error map between the generated images and the ground truth(GT) images. Red boxes demonstrate that our method generates more realistic images with clearer textures and richer structural details.}
\label{fig3}
\end{figure*}

\section{Experiments}
\subsection{Settings}
\subsubsection{Datasets} We conduct all our experiments on the BRATS 2020 dataset
(\url{https://www.med.upenn.edu/cbica/brats2020/data.html})\cite{bakas2017advancing}, which provides brain tumor images of four modalities: Flair, T1, T1ce and T2. All images have been manually segmented by experienced raters. Annotations include the GD-enhancing tumor (ET), peritumoral edema (ED), as well as the necrotic and non-enhancing tumor core (NCR/NET)\cite{menze2014multimodal, bakas2018identifying}. The tumors are classified into three distinct regions according to clinical references: Whole Tumor (WT = ET + ED + NCR/NET), Tumor Core (ET + NCR/NET), and Enhancing Tumor (ET). We calculated masks corresponding to these three areas based on the annotations to guide the brain image translation. 180 samples are randomly selected for training, 30 for validation and 159 for testing in the translation task. For each sample, we select images from only one randomly chosen modality. We trained three sets of models using the aforementioned three types of tumor masks, enabling translation between any two of the mentioned four modalities. To further verify the effectiveness of the model, we feed the fully multi-modality images generated by different methods into nnU-Net\cite{isensee2021nnu} to compare their segmentation effectiveness as a downstream segmentation task. 120 samples are selected for training and 69 for testing. It is worth noting that none of the samples used for the downstream segmentation task are included in the training set for the translation task, avoiding the resulting metric being biased positively towards our model. All the images are resized to $128\times128$ pixels.

\begin{table*}[htb]
\scriptsize
\centering
\caption{Quantitative evaluation of the generated images. We report the mean value for translation between any two modalities here. The symbol ↑ denotes higher is better while the symbol ↓ denotes lower is better. UTAD-NET(WT) denotes our model guided by the Whole Tumor mask, and the same applies to the others.}
\begin{tabular}{ccccccc}
\hline
\multicolumn{2}{c|}{Method}                                                        & SSIM↑ & PSNR↑ & local SSIM↑ & local PSNR↑ & LPIPS↓ \\ \hline
\multicolumn{1}{c|}{\multirow{5}{*}{Baseline}} & \multicolumn{1}{c|}{StarGAN\cite{choi2018stargan}}     & 0.8477±0.0005 & 28.02±1.30 & 0.5886±0.0008       & 18.48±1.40       & 0.0930±0.0228   \\
\multicolumn{1}{c|}{}                          & \multicolumn{1}{c|}{DRIT++\cite{lee2020drit++}}      & 0.8615±0.0008 & 30.24±1.20 & 0.6095±0.0009       & 19.16±1.26       & 0.0802±0.0276   \\
\multicolumn{1}{c|}{}                          & \multicolumn{1}{c|}{TarGAN\cite{chen2021targan}}      & 0.8588±0.0006 & 29.66±1.24 & 0.6037±0.0010       & 18.78±1.31       & 0.0860±0.0235   \\
\multicolumn{1}{c|}{}                          & \multicolumn{1}{c|}{ReMIC\cite{shen2020multi}}       & 0.8459±0.0007 & 29.31±1.41 & 0.6010±0.0013       & 18.69±1.14       & 0.0874±0.0262   \\ 
\multicolumn{1}{c|}{}                          & \multicolumn{1}{c|}{ResViT\cite{dalmaz2022resvit}}       & 0.8735±0.0005 & 30.45±1.37 & 0.6068±0.0009       & 19.33±1.20       & 0.0702±0.0236   \\
\hline
\multicolumn{1}{c|}{\multirow{3}{*}{Teacher}} & \multicolumn{1}{c|}{UTAD-Net(WT)} & 0.8757±0.0004 & 31.57±1.14 & 0.6195±0.0008       & \textbf{20.72±1.33}       & 0.0655±0.0214   \\
\multicolumn{1}{c|}{}                          & \multicolumn{1}{c|}{UTAD-Net(TC)} & \textbf{0.8767±0.0005} & \textbf{31.89±1.29} & \textbf{0.6199±0.0007}       & 20.44±1.22       & \textbf{0.0647±0.0249}   \\
\multicolumn{1}{c|}{}                          & \multicolumn{1}{c|}{UTAD-Net(ET)} & 0.8751±0.0006 & 31.61±1.42 & 0.6166±0.0008       & 20.39±1.16       & 0.0690±0.0243   \\ \hline
\multicolumn{1}{c|}{\multirow{3}{*}{Student}}  & \multicolumn{1}{c|}{UTAD-Net(WT)} & 0.8742±0.0006 & \textbf{31.48±1.26} & 0.6147±0.0009       & \textbf{20.21±1.23}       & 0.0664±0.0230   \\
\multicolumn{1}{c|}{}                          & \multicolumn{1}{c|}{UTAD-Net(TC)} & \textbf{0.8753±0.0004} & 31.34±1.30 & \textbf{0.6156±0.0012}       & 19.96±1.26       & \textbf{0.0661±0.0255}   \\
\multicolumn{1}{c|}{}                          & \multicolumn{1}{c|}{UTAD-Net(ET)} & 0.8723±0.0005 & 30.88±1.15 & 0.6143±0.0008       & 19.93±1.17       & 0.0716±0.0268   \\ \hline
\multicolumn{1}{c|}{\multirow{2}{*}{Ablation}}  & \multicolumn{1}{c|}{UTAD-Net(zeros map)} & 0.8535±0.0007 & 29.26±1.48 & 0.6013±0.0005      & 18.75±1.59      & 0.0869±0.0264  \\
\multicolumn{1}{c|}{}                          & \multicolumn{1}{c|}{UTAD-Net(random map)} & 0.8509±0.0007 & 28.47±1.40 & 0.5994±0.0009      & 18.26±1.31      & 0.0884±0.0281  \\ \hline
\end{tabular}
\label{Tab1}
\end{table*}

\subsubsection{Evaluation metrics} For the translation task, we use structural similarity index measure (SSIM), peak-signal-noise ratio (PSNR)\cite{hore2010image} and learned perceptual image patch simila-rity (LPIPS)\cite{zhang2018unreasonable} to measure the similarity between the generated images and ground truth. For the downstream segmentation task, we use Dice similarity coefficient (DSC)\cite{dice1945measures}, average symmetric surface distance (ASSD),  95th percentile of Hausdorff distance (HD95) to measure the integrity of the predicted pseudo masks generated by nnU-Net\cite{isensee2021nnu} which is an acknowledged state-of-the-art medical image segmentation model. 

\subsubsection{Baselines} We compare our translation results with StarGAN
\cite{choi2018stargan}, DRIT++\cite{lee2020drit++}, Targan\cite{chen2021targan}, ReMIC\cite{shen2020multi} and ResViT\cite{dalmaz2022resvit}. StarGAN proposes using a unified model to translate images to arbitrary modalities. DRIT++ disentangles an image into the content code and the attribute code during the training and generates images using the content code extracted from the input images and the attribute code sampled from the standard normal distribution. TarGAN utilizes an extra shape controller to alleviate the problem of image deformation in the target area. ReMIC learns shared content and domain-specific style encoding across multiple domains. ResViT proposes a unified synthesis model with a transformer-based generator for medical image synthesis. Note that we implement unsupervised versions for ReMIC and ResViT to enable a fair comparison. 

\begin{table*}[!t]
\scriptsize
\centering
\caption{The quantitative evaluation for downstream segmentation task conducted by nnU-Net.}
\begin{tabular}{cc|ccc|ccc|ccc}
\hline
\multicolumn{2}{c|}{\multirow{2}{*}{Method}}                          & \multicolumn{3}{c|}{DSC}                    & \multicolumn{3}{c|}{ASSD(mm)}                 & \multicolumn{3}{c}{HD95(mm)}                  \\ \cline{3-11} 
\multicolumn{2}{c|}{}                                                 & WT             & TC             & ET             & WT            & TC            & ET            & WT            & TC            & ET            \\ \hline
\multicolumn{1}{c|}{\multirow{6}{*}{Baseline}} & SingleGT             & 0.8012±0.0217          & 0.6944±0.0256          & 0.4775±0.0273          & 1.69±0.52          & 2.44±0.65          & 2.92±0.61          & 5.69±1.48          & 6.72±1.65          & 7.73±1.59          \\
\multicolumn{1}{c|}{}                          & StarGAN\cite{choi2018stargan}              & 0.7520±0.0243          & 0.6357±0.0284          & 0.4018±0.0310          & 3.77±0.72          & 4.01±0.76          & 3.70±0.88          & 10.57±3.02         & 10.89±3.16         & 11.48±3.28         \\
\multicolumn{1}{c|}{}                          & DRIT++\cite{lee2020drit++}               & 0.7867±0.0226          & 0.6749±0.0271          & 0.4403±0.0263          & 2.53±0.56          & 2.62±0.84          & 2.81±0.63          & 7.79±2.54          & 8.12±2.10          & 8.63±2.26          \\
\multicolumn{1}{c|}{}                          & TarGAN\cite{chen2021targan}               & 0.7782±0.0283          & 0.6670±0.0265          & 0.4361±0.0301          & 2.51±0.75          & 2.89±0.72          & 2.96±0.80          & 7.93±2.67          & 8.24±2.41          & 8.81±2.79          \\
\multicolumn{1}{c|}{}                          & ReMIC\cite{shen2020multi}                & 0.7737±0.0294          & 0.6618±0.0322          & 0.4328±0.0306          & 2.63±0.63          & 2.95±0.68          & 2.78±0.81          & 8.03±2.60          & 8.39±2.23          & 8.75±2.59          \\
\multicolumn{1}{c|}{}                          & ResViT\cite{dalmaz2022resvit}                & 0.7983±0.0233          & 0.7037±0.0270          & 0.4756±0.0286          & 2.83±0.69          & 2.29±0.78          & 2.68±0.74          & 7.43±2.49          & 7.03±2.28          & 7.56±2.64          \\ \hline
\multicolumn{1}{c|}{\multirow{3}{*}{Teacher}}  & UTAD-Net(WT)          & \textbf{0.8190±0.0224} & 0.6987±0.0243          & 0.4573±0.0290          & \textbf{1.53±0.56} & 2.14±0.60          & 2.32±0.78          & \textbf{3.89±1.79} & 6.17±1.50          & 7.80±2.46          \\
\multicolumn{1}{c|}{}                          & UTAD-Net(TC)          & 0.8040±0.0256          & \textbf{0.7432±0.0284} & 0.5176±0.0293          & 1.79±0.63          & \textbf{1.81±0.70} & 1.75±0.74          & 6.09±1.84          & \textbf{5.78±2.06} & 5.71±2.36          \\
\multicolumn{1}{c|}{}                          & UTAD-Net(ET)          & 0.8004±0.0263          & 0.7097±0.0305          & \textbf{0.5452±0.0303} & 1.92±0.65          & 2.17±0.72          & \textbf{1.68±0.79} & 6.50±2.42          & 6.40±2.52          & \textbf{5.46±2.27} \\ \hline
\multicolumn{1}{c|}{\multirow{3}{*}{Student}}  & UTAD-Net(WT)          & \textbf{0.8084±0.0258} & 0.6937±0.0262          & 0.4483±0.0294          & \textbf{1.62±0.60} & 2.56±0.68          & 2.52±0.71          & \textbf{5.17±2.15} & 7.10±2.36          & 8.05±2.63          \\
\multicolumn{1}{c|}{}                          & UTAD-Net(TC)          & 0.7992±0.0241          & \textbf{0.7184±0.0285} & 0.5011±0.0321          & 1.82±0.60          & \textbf{2.05±0.63} & 2.49±0.69          & 6.17±2.48          & \textbf{6.49±2.35} & 6.76±2.65          \\
\multicolumn{1}{c|}{}                          & UTAD-Net(ET)          & 0.7953±0.0262          & 0.6840±0.0246          & \textbf{0.5255±0.0255} & 2.13±0.67          & 2.59±0.66          & \textbf{2.42±0.75} & 6.62±2.62          & 7.96±2.58          & \textbf{6.26±2.56} \\ \hline
\multicolumn{1}{c|}{\multirow{2}{*}{Ablation}} & UTAD-Net(zeros map)  & 0.7772±0.0265          & 0.6652±0.0294          & 0.4308±0.0283          & 2.67±0.62          & 2.91±0.76          & 2.93±0.75          & 8.01±2.76          & 8.44±2.66          & 8.92±2.68          \\
\multicolumn{1}{c|}{}                          & UTAD-Net(random map) & 0.7756±0.0248          & 0.6631±0.0315          & 0.4306±0.0287          & 2.75±0.65         & 2.96±0.73          & 2.86±0.86          & 8.10±2.80          & 8.52±2.78          & 8.84±2.77          \\ \hline
\end{tabular}
\label{Tab2}
\end{table*}

\subsubsection{Implementation details} We implement PatchGAN\cite{isola2017image} as the backbone for both the global discriminator and local discriminator, and U-net\cite{ronneberger2015u} as the backbone for the generator. We train the teacher network for $100$ epochs with a learning rate of $10^{-4}$ for both the generator and the discriminators during the first $50$ epochs and then linearly decay the learning rate to $10^{-6}$ at the final epoch. Then we train the student network in the same way. Adam\cite{kingma2014adam} optimizer is used with momentum parameters ${\beta}_1=0.9$ and ${\beta}_2=0.999$. We also adopt data augmentation and normalization for the training samples. All the experiments are conducted on PyTorch 1.8.1 with NVIDIA RTX 3090(24G).

\subsection{Translation results}

\subsubsection{Qualitative evaluation} 
\Cref{fig3} shows the qualitative results of our model and the other baselines. Our method generates more realistic images with clearer textures and richer structural details.

\begin{figure*}[t]
    \centering
    \includegraphics[width=1.0\linewidth]{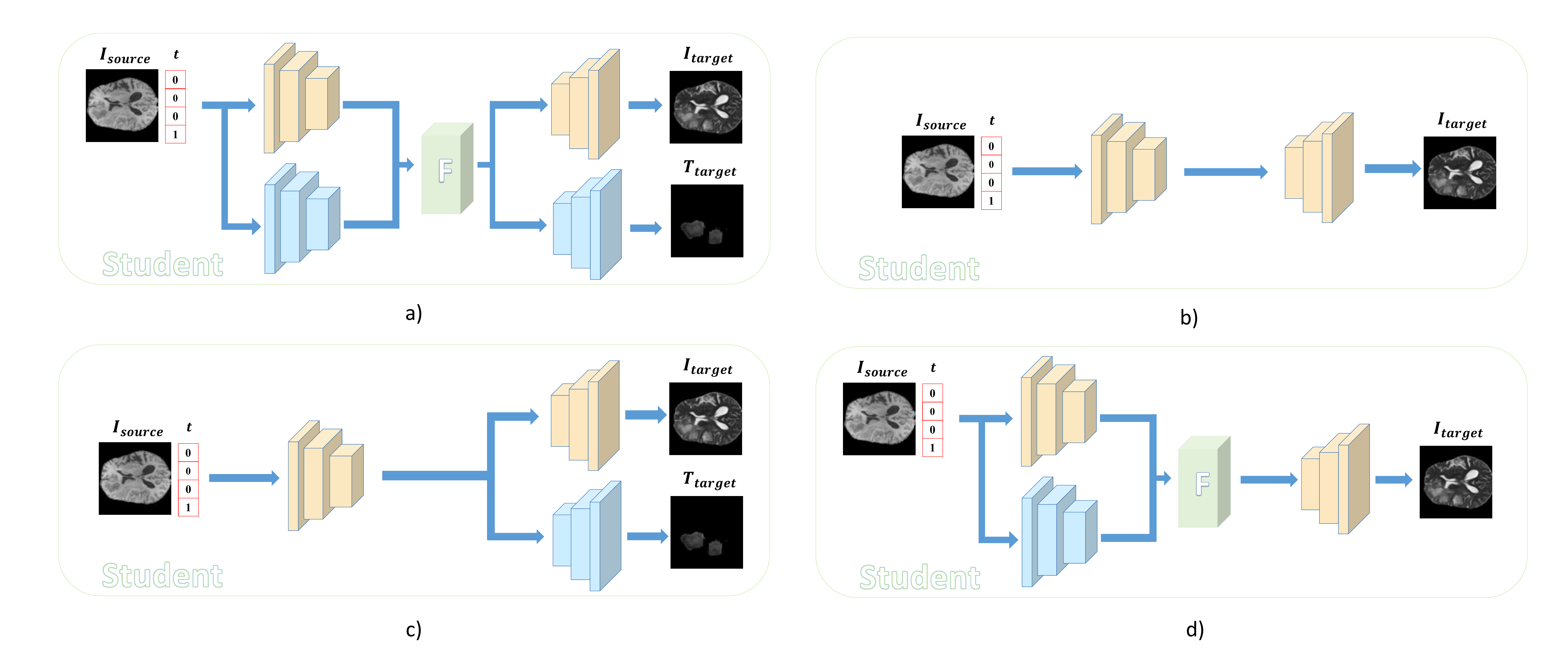}
    \caption{Four architecture schemes of the student network.a) Complete global branch and local branch (UTAD-Net). b) Only global branch. c) Without local tumor encoder. d) Without local tumor decoder. }
    \label{fig4}
\end{figure*}

\subsubsection{Quantitative evaluation} As shown in \Cref{Tab1}, the teacher networks of our models guided by different masks significantly outperform the baseline on various metrics, and the performance of the student network is slightly worse than the teacher network but still better than all baselines. The value is the average of all the possible mappings. In addition to SSIM, PSNR and LPIPS, we use the smallest rectangle to frame areas of Whole Tumor for every generated image and calculate SSIM and PSNR of the framed areas with their corresponding ground truth, which is denoted as local SSIM and local PSNR. These two metrics can measure the translation effect of the tumor areas. UTAD-Net performs better on the two metrics, which suggests that our method generates more realistic tumor areas. 

\subsection{Downstream segmentation results}
Given an image from an arbitrary modality, we translate it to the other three modalities by our method and all the baselines respectively. Then we use nnU-Net to compare the segmentation effectiveness of the fully multi-modality images generated by each of the above methods. As shown in \Cref{Tab2}, the teacher network achieves optimal results, which suggests that guiding with the tumor masks leads to more realistic tumor areas. Note that the SingleGT refers to randomly selecting one modality of the ground truth for each sample in the training set for segmentation. Although the student network's performance is slightly lower than that of the teacher network, it still outperforms all baselines and singleGT on the corresponding downstream segmentation tasks, indicating that the images generated by our model have practical clinical significance.

\begin{figure}
    \centering
    \includegraphics[width=1\linewidth]{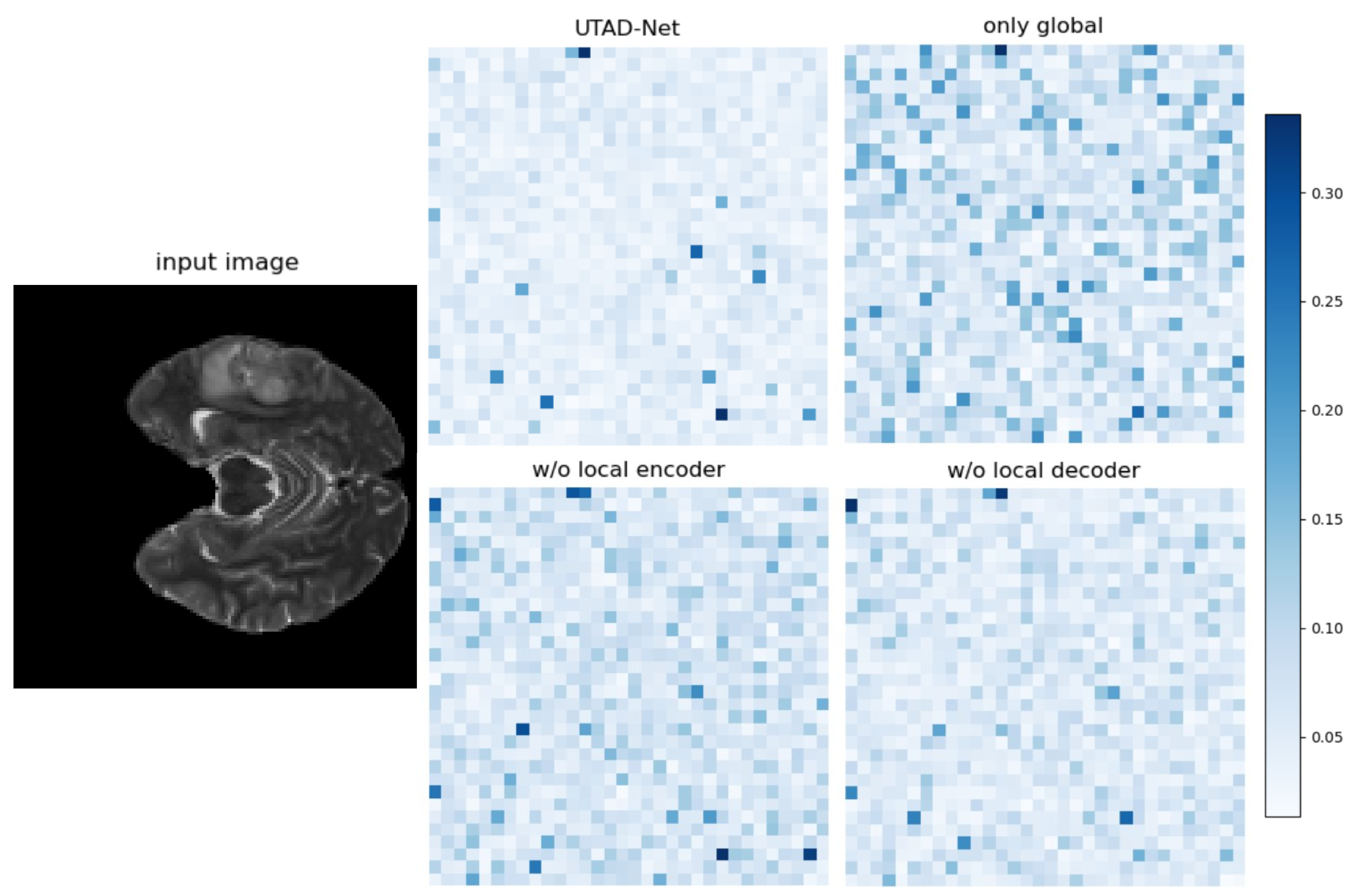}
    \caption{The feature error map of the teacher network and the student network. Each pixel represents a channel.}
    \label{fig5}
\end{figure}

\subsection{Ablation study}
In this section, we conduct two parts of the ablation study, one to validate the effectiveness of the local branch in the teacher network, and the other one to confirm the efficacy of the student network structure.
\subsubsection{Effectiveness of the local branch in the teacher network}
We conduct an ablation study to validate the effectiveness of the local branch which is guided by the mask. We replace the tumor mask in the teacher network with two settings: (a) a map with all zero values. (b) a random map that each value is either zero or one. As shown in \Cref{Tab1} and \Cref{Tab2}, the performance of UTAD-Net with a map of zeros (i.e., translation without the guidance of a tumor mask) is degenerated in terms of image quality and downstream segmentation. UTAD-Net with a random map performs even worse, because the noise images without semantic information have a negative effect on the model's performance.

\subsubsection{The structure of the student network}
The purpose of the student network is to perceive the tumor areas like the teacher network without tumor masks. In order to learn the features extracted by the teacher network from the tumor areas better, we designed the following four different architecture schemes of the student network for comparison, as illustrated in \Cref{fig4}. We compute the error between the features obtained from the fusion block of the teacher network and the corresponding layer features of student network, averaging across different pixels to obtain the feature error map along the channel dimension. Note that for student networks in structures a) and d), we select the features obtained from the fusion block, whereas for structures b) and c), we choose the features from the last layer of the encoder. The error map is shown in \Cref{fig5}. The absence of either the local encoder or local decoder structure results in a reduction in the accuracy of the student network. The student network with only the global branch performs the worst, while maintaining the same structure as the teacher network yields the most similar features. We use Mean Absolute Error (MAE) and Mean Squared Error (MSE) to assess the similarity of features between the teacher network and the student network, where lower values indicate higher similarity. The quantified results are presented in \Cref{Tab3}, demonstrating that the structure combining both global and local branches ultimately yields the best performance.

\begin{table}[t]
\centering
\caption{The quantitative evaluation of the feature error for the teacher network and the four different architecture schemes of the student network.}
\begin{tabular}{c|c|c}
\hline
\textbf{}         & MAE             & MSE             \\ \hline
UTAD-Net           & \textbf{0.0500} & \textbf{0.0124} \\
only global       & 0.2514          & 0.2229          \\
w/o local encoder & 0.1912          & 0.1588          \\
w/o local decoder & 0.1790          & 0.1423          \\ \hline
\end{tabular}
\label{Tab3}
\end{table}

\section{Conclusion}
We propose UTAD-Net to translate brain tumor images, in order to capture tumor areas precisely and generate more realistic images between any two modalities. With the guidance of the tumor masks, the teacher network can focus on the brain tumor areas and alleviate the problem of deformation in the generated images. The student network gains the ability to perceive the tumor areas and generate target images without masks through distillation. Experiments demonstrate that our model achieves superior translation results and generates more realistic images with practical clinical significance, which results in improved downstream segmentation.

\section{Acknowledgments}
This work is supported in part by the Guangdong Provincial Natural Science Foundation (2023A1515011431), the Guangzhou Science and Technology Planning Project (202201010092), the National Natural Science Foundation of China (72074105), NSF-1850492 and NSF-2045804.

\bibliographystyle{IEEEtran}
\bibliography{IEEEabrv, ijcnn_2024}

\end{document}